\documentclass[12pt]{article}
\usepackage{amsfonts,amsmath,epsfig,graphicx,amssymb}
\usepackage{amsmath,epsfig,graphicx,amssymb}
\usepackage[margin=1in]{geometry}
\usepackage{amsmath}
\usepackage{amsfonts}
\usepackage{amssymb}
\usepackage{makeidx}
\usepackage{graphicx}
\usepackage{float}
\newcommand{\beq}{\begin{equation}}
\newcommand{\eeq}{\end{equation}}
\newcommand{\ben}{\begin{eqnarray}}
\newcommand{\een}{\end{eqnarray}}
\date{}
\begin{document}
\title{{Derivation of a Schr\"{o}dinger Equation for Single Neurons Through Stochastic Neural Dynamics.}}
\author{Partha Ghose \footnote{partha.ghose@gmail.com}\\Tagore Centre for Natural Sciences and Philosophy,\\ Rabindra Tirtha, New Town, Kolkata 700156, India} 
\maketitle
\begin{abstract}
Despite the prevalent view that quantum mechanics is irrelevant to macroscopic biological systems because of inherent noise and decoherence, this paper demonstrates that the electrical noise (Brownian motion) in neuron membranes gives rise to an `emergent' Schr\"{o}dinger equation involving a new neuronal constant $\hat{\hbar}$, fundamentally challenging the standard view of neuronal behaviour. This result could provide new insights into the underlying mechanisms of brain function, thus challenging existing paradigms in both quantum physics and neuroscience. A possible empirical test of this emergent quantum behaviour would be to look for quantum fluctuations in subthreshold neural oscillations. 
\end{abstract}

Quantum mechanics is widely regarded as the most fundamental theory of nature, suggesting its applicability to even brain functions. However, the prevailing view is that quantum coherence is lost at the neuronal level due to the large size and complex environment of the neurons, a process known as `decoherence' \cite{teg, koch}. As a result, both physicists and neuroscientists have traditionally dismissed the relevance of quantum mechanics in neuronal processes. Unlike previous works that primarily focus on classical interpretations of stochastic processes in neurons, our approach explores the quantum-like dynamics that `emerges' directly from such processes.

Building on Nelson's method of deriving the Schr\"{o}dinger equation from underlying classical stochastic processes \cite{nel1} and its further developments \cite{ nel2, nel3, com, gu, far, pet, wang}, this paper demonstrates how stochastic neural dynamics can exhibit emergent quantum-like behaviour exhibiting coherence, stability and structure, a surprising example of `order-from-disorder' \cite{schr}. Such a perspective offers a new framework for understanding brain function in terms of quantum-like processes.
\begin{figure}[H]
\centering
{\includegraphics[scale=0.3]{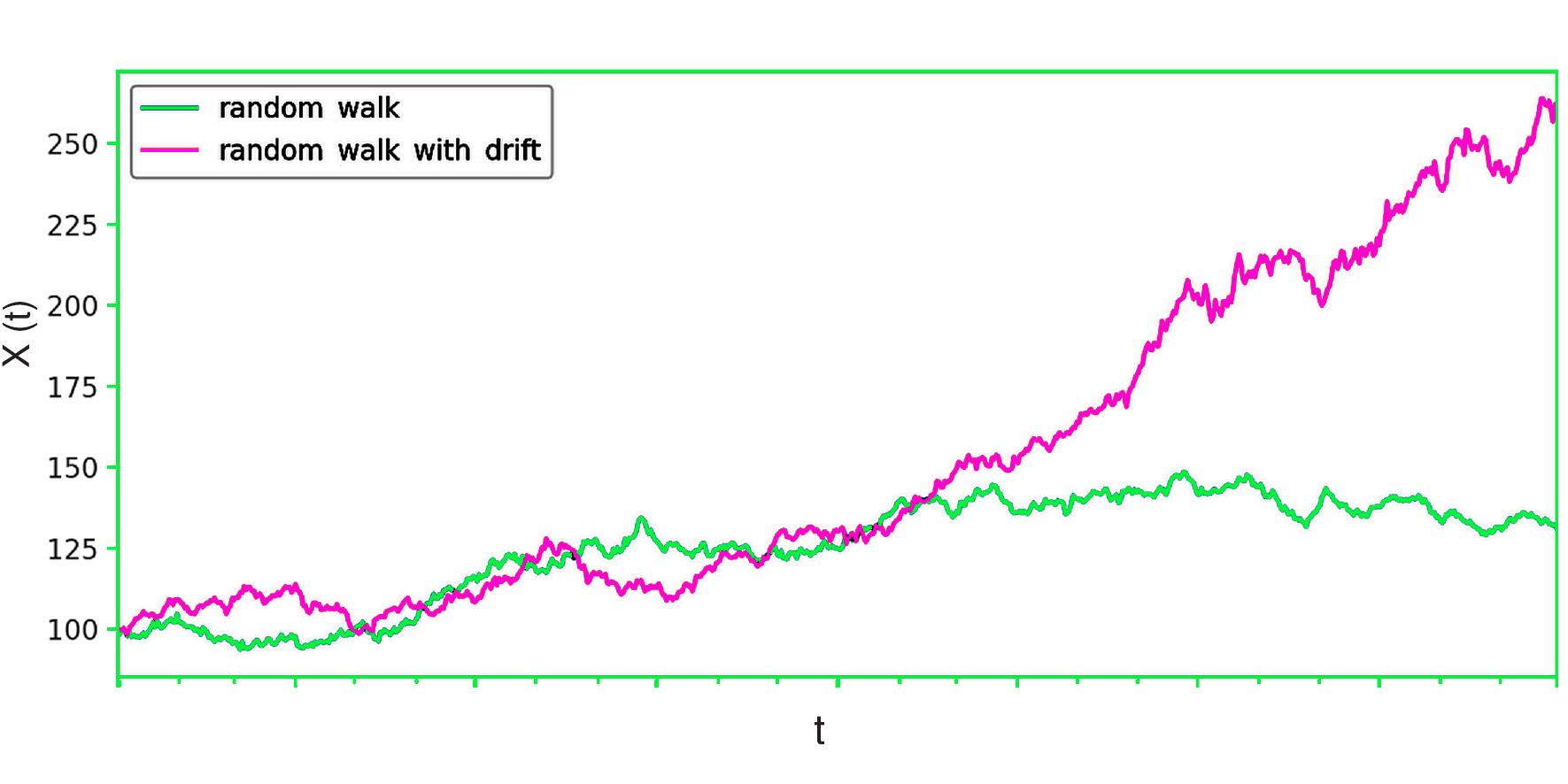}}
\caption{\label{Figure 1}{\footnotesize Random walk with and without drift. The value $x(t)$ of the random variable $X(t)$ at time $t$ equals the last period's value plus a constant (a drift) and a white noise. The drift can be in the direction of the threshold (forward) or away from it (backward).}}
\end{figure}
\begin{figure}[H]
\centering
{\includegraphics[scale=0.7]{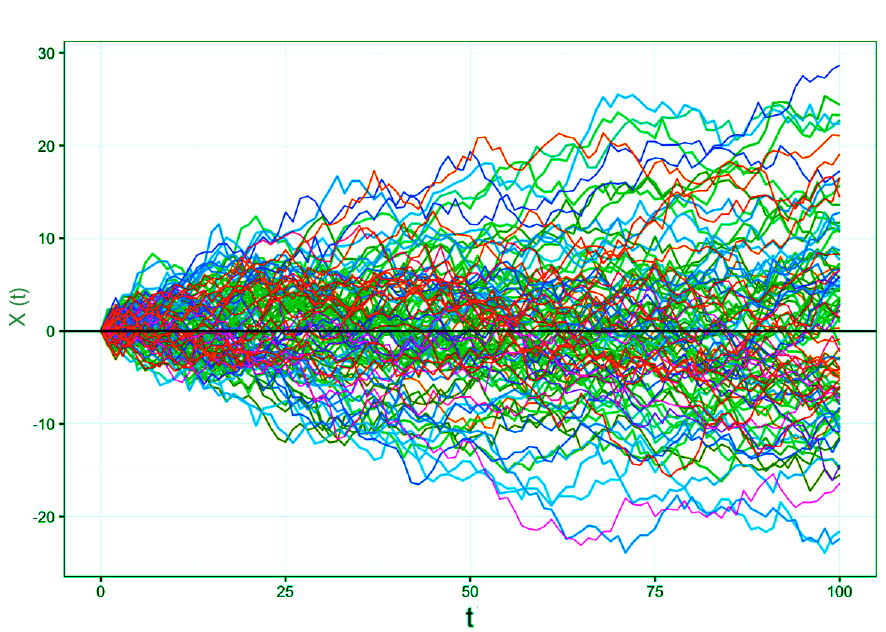}}
\caption{\label{Figure 2}{\footnotesize Wiener process or gaussian random walk (Brownian motion) showing diffusion, i.e. increase of variance with time.}}
\end{figure}

A substantial body of empirical and theoretical research has already established that neurons and neural networks exhibit stochastic behaviour \cite{car, faisal}. A stochastic process is a sequence of random variables whose values change over time in an uncertain way so that one knows only the distribution of the possible values at any point in time. It will be useful to introduce a few more related technical terms. A Markov stochastic process is a particular type of stochastic process where only the current value of a variable is relevant for predicting the future movement, not past values.  A random walk is the stochastic process formed by successive summation of independent, identically distributed random variables. It is one of the most basic and well-studied topics in probability theory. A Wiener process, also called Brownian motion, is a Markov process which is essentially a series of normally distributed random variables such that for later times the variances of these normally distributed random variables increase, a process called `diffusion'. 
\begin{figure}[H]
\centering
{\includegraphics[scale=0.6]{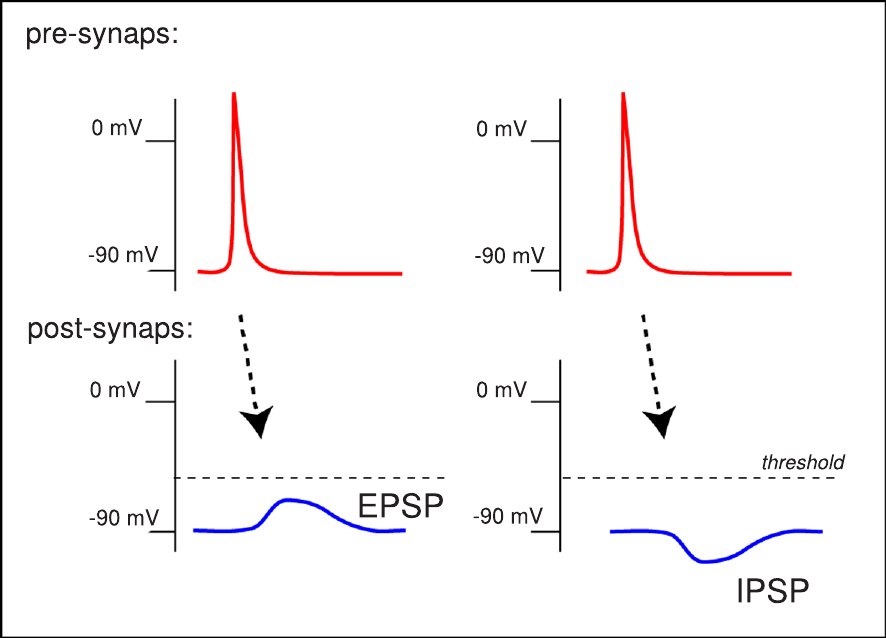}}
\caption{\label{Figure 3}{\footnotesize Schematic representation of a neuron's state evolving under stochastic dynamics. The figure illustrates the membrane potential's Brownian motion, with the forward and backward processes corresponding to Excitatory (EPSP) and Inhibitory (IPSP) Post-Synaptic Potentials, respectively.}}
\end{figure}

Notably, random walk and diffusion models have been proposed to describe neuronal dynamics and spike activity. For instance, the electrical state of polarization in the somatic and dendritic membrane can be modeled as a `state point' executing a random walk, influenced by excitatory and inhibitory post-synaptic potentials (EPSPs and IPSPs) \cite{gers, clay}. Each incoming elemental EPSP (Excitatory Post Synaptic Potential) moves the state point one unit toward the threshold of neuron firing, and each incoming elemental IPSP (Inhibitory Post Synaptic Potential) moves the state point one unit away from the threshold. If the average rate of incoming elemental EPSP and elemental IPSP are the same, there is an equal probability at any time that the state point moves either a unit toward or a unit away from the threshold, i.e. there is no ``bias toward'' either input. Immediately after the state point has attained the threshold and caused the production of an action potential, it returns to the resting potential, only to begin its random walk again. Such a model would be a simple random walk model. 

In reality, however, these two rates may sometimes be different, and in realistic physiological models it would be far more reasonable to assume that that there is some excess of either EPSP or IPSP input. In this case, the probability for the state point to move one unit toward the threshold will be different from the probability for it to move away from the threshold. Considered as a diffusion process, the difference between these probabilities can be considered a``drift velocity'', either toward or away from the threshold. 

Our aim being to demonstrate that the inherent stochastic nature of neural processes, as specifically implied by `random walk with drift' models, can generate a Schr\"{o}dinger-like equation, we will treat the membrane potential $X(t)$ as a random variable and establish a connection between classical diffusion processes and emergent quantum-like behavior in neurons.

Following Nelson's approach, let us then consider the following model. The “state point” $X(t)$ of a canonical neuron (consisting of a {\em dendritic arbor}, the {\em soma} and the {\em axon}) represents the membrane potential. As time passes and the electrical state of the membrane varies randomly, the state point moves back and forth along a straight line, executing Brownian motion without friction. Then $X(t)$ follows the stochastic differential equation (SDE)
\beq
dX(t) = b_f(X(t), t)dt + \sigma dW_f(t).
\eeq
Here, $b_f$ represents the forward drift velocity (caused by EPSPs), $\sigma$ is the square root of the diffusion coefficient, and $dWf(t)$ is a forward Wiener process. Since these processes are conservative, backward processes (caused by IPSP) also exist.
The SDE for such backward processes is
\beq
dX(t) = b_b(X(t), t)dt + \sigma dW_b(t)
\eeq
where $dW_b(t)$ is the backward Wiener process. The diffusion coefficient $\sigma^2$ is determined by the physiological characteristics of the membrane.

The solutions $X(t)$ of the stochastic equations are known to be continuous at all state points but nowhere differentiable. Hence, Nelson suggested the following average forward and backward differentials which we adopt:
\ben
D_fX(t) &=& \lim_{\Delta t\rightarrow 0} E_t\left[\frac{X(t + \Delta t) - X(t)}{\Delta t}\right],\\
D_bX(t) &=& \lim_{\Delta t\rightarrow 0} E_t\left[\frac{X(t) - X(t - \Delta t)}{\Delta t}\right]
\een
where $E_t$ denotes the expectation conditional on $X(t) = x$. For differentiable curves $D_fX(t) = D_bX(t) = \dot{x} = v(t)$, the `velocity' of the state point. 

To bridge this stochastic description with quantum mechanics, we now proceed to derive a Schr\"{o}dinger-like equation from the underlying diffusion (Wiener) process. Following Nelson, let us postulate Newton's law for the stochastic acceleration,
\beq
m a (X(t)) = m\frac{1}{2}(D_fD_b + D_bD_f)X(t) = F(X(t))
\eeq
where $m$ denotes state inertia, the ability of biological systems to keep a functional state at rest or in activity and is an active process of resistance to change in state.
It follows that the drift coefficients in the forward and backward equations are given by
\beq
D_f(X(t)) = b_f(X(t), t),\,\,\,D_b(X(t)) = b_b(X(t), t)
\eeq
This amounts to a complete description of the motion, as in classical mechanics. 

As is well known, the forward and backward SDEs lead to two Fokker-Planck equations for the probability density $\rho(x,t)$ of the random variable executing Brownian motion:
\ben
\frac{\partial}{\partial t}\rho(x,t) &=& -\frac{\partial}{\partial x}\left[b_f(x,t)\rho(x,t)\right] + \frac{\sigma^2}{2}\frac{\partial^2}{\partial x^2}\rho(x,t),\\
\frac{\partial}{\partial t}\rho(x,t) &=& -\frac{\partial}{\partial x}\left[b_b(x,t)\rho(x,t)\right] - \frac{\sigma^2}{2}\frac{\partial^2}{\partial x^2}\rho(x,t)
\een
Adding these equations results in the continuity equation
\beq
\frac{\partial}{\partial t}\rho(x,t) + \frac{\partial}{\partial x}[v(x,t) \rho(x,t)] = 0 \label{cont}
\eeq
with the {\em current velocity} defined by $v(x,t) = (b_f(x,t) + b_b(x,t))/2$. This displays the role of the current velocity in maintaining the probability distribution:

The difference of the forward and backward drifts $u(x,t) = (b_f(x,t) - b_b(x,t))/2$ is defined as the {\em osmotic velocity}. Subtracting the two Fokker-Planck equations results in
\beq
u(x,t) = \frac{\sigma}{2}\frac{\partial}{\partial x}\ln[\rho(x,t)] = \frac{\sigma}{2}\frac{\partial_x \rho}{\rho} = \sigma\frac{\partial R}{\partial x}\label{os}  
\eeq
where $\ln\rho(x,t) = 2R(x,t)$. The coupled forward-backward stochastic differential equations for the position process can thus be written as
\ben
dX(t) &=& \left(v(X(t),t) + u(X(t), t)\right) + \sigma dW_f(t),\label{sde1}\\
dX(t) &=& \left(v(X(t),t) - u(X(t), t)\right) + \sigma dW_b(t).\label{sde2}
\een
It follows from this that the current velocity is curl-free and can be written as
\beq
v(x,t) = \frac{1}{m}\frac{\partial }{\partial x}S(x,t) \label{v}
\eeq
where $S(x,t)$ is a scalar function which can be identified with the action.  

Now, following Guerra and Morato \cite{gu}, let us introduce the Lagrangian field
\beq
{{\cal{L}}} = \frac{1}{2}m (v^2 - u^2)(x,t) - V(x)
\eeq
where $V(x)$ is the electrostatic potential, from which the action $S(x,t)$ can be constructed. It can then be shown, using the variational principle, that the main features of Nelson's stochastic mechanics including eqns (\ref{os}) and (\ref{v}) can be derived from such an action.

Using stochastic control theory and the current velocity as the control, Guerra and Morato showed that the following differential equations for the functions $R$ and $S$ extremize the action:
\ben
\frac{\partial S}{\partial t} + \frac{1}{2m}\left(\frac{\partial S}{\partial x}\right)^2 + V + V_Q &=& 0,\,\,\, V_Q = -\frac{m\sigma^2}{2}\left[\left(\frac{\partial R}{\partial x}\right)^2 + \frac{\partial^2 R}{\partial x^2}\right],\label{shj}\\
\frac{\partial R}{\partial t} + \frac{1}{2m} \left(R\frac{\partial^2 S}{\partial x^2} + 2\frac{\partial R}{\partial x}\frac{\partial S}{\partial x}\right) &=& 0 \label{con}.
\een 
The first equation is the Hamilton-Jacobi-Bellman equation, i.e., the Hamilton-Jacobi equation with an additional stochastic term  $V_Q$ which takes the form
\beq
V_Q = -\frac{m\sigma^2}{4}\left[\frac{\partial_x^2\rho}{\rho} -\frac{(\partial_x \rho)^2}{2\rho^2}\right]
\eeq
in terms of $\rho = e^{2R}$. It is the analog of the Bohm quantum potential \cite{bohm}. The second equation can also be written in terms of $\rho$ as
\beq
\frac{\partial \rho}{\partial t} + \partial_x\left[\rho \frac{\partial_x S}{m}\right] = 0
\eeq
which, using eqn (\ref{v}) for the current velocity, is a continuity equation. These two coupled partial differential equations determine the stochastic process. These equations can be derived from the Schr\"{o}dinger-like equation
\beq
im\sigma\frac{\partial}{\partial t}\psi(x,t) = \left(-\frac{m\sigma^2}{2}\partial_x^2 + V(x) \right)\psi(x,t) \label{sc}
\eeq
by putting $\psi = \exp (R + iS/m\sigma) = \sqrt{\rho}\exp (iS/m\sigma)$ and separating the real and imaginary parts \cite{gu, bohm}. The argument $x$ in the wave function $\psi(x,t)$ represents the `state point' of the neuron membrane and plays the role of the spatial coordinate and the coefficient $\sigma$ plays the role of the factor $\hbar/m$ in standard quantum mechanics. 

It should be borne in mind that the stochastic processes which occur in neural membranes are different from those that give rise to standard quantum mechanics which is responsible for the stability and structure of the atoms and molecules which constitute the neurons. It would therefore be convenient to introduce a new universal constant $\hat{\hbar} = m\sigma$ for neuronal media and rewrite the above equation in the form
\beq
i\hat{\hbar}\frac{\partial}{\partial t}\psi(x,t) = \left(-\frac{\hat{\hbar}^2}{2m}\partial_x^2 + V(x) \right)\psi(x,t).\label{sc2}
\eeq
and treat it as the Schr\"{o}dinger equation for canonical neurons. Like the standard Schr\"{o}dinger equation in quantum mechanics, this equation will also ensure a novel level of stability and structure, i.e. stable order in the stochastic world of neurons.

The wave function $\psi$ describes the Markov process completely:
\ben
\rho &=& |\psi|^2,\\
u &=& \sigma\partial_x \Re \ln \psi,\\
v &=& \sigma\partial_x \Im \ln \psi.
\een
This is the `Nelson map'. It maps the probability distribution function and the current and osmotic velocities in the neurons to a wave function. In other words, it associates a diffusion process in the neurons to every solution of the Schr\"{o}odinger-like equation (\ref{sc2}).

\begin{figure}[H]
\centering
{\includegraphics[scale=0.6]{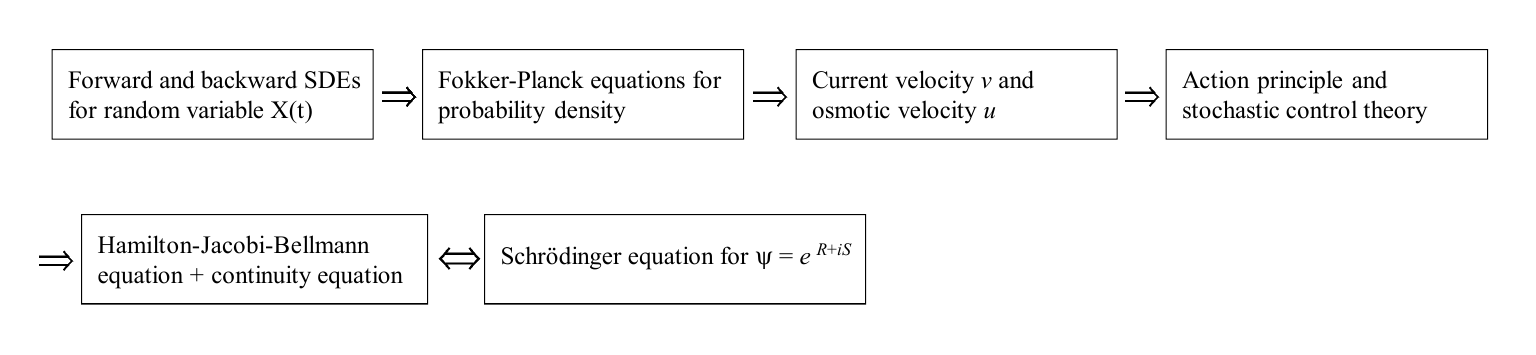}}
\caption{\label{Figure 4}{\footnotesize Flow chart depicting the main mathematical steps used in deriving the Schr\"{o}dinger equation from stochastic mechanics.}}
\end{figure}

A possible experimental validation of the ideas proposed in this paper would be to look for `quantum-like fluctuations' in subthreshold neural oscillations. Subthreshold oscillations are membrane oscillations that do not directly trigger an action potential because they do not reach the necessary threshold for firing. However, they may facilitate sensory signal processing \cite{sch}. Unlike oscillations that trigger action potentials and can be modelled by the charging and discharge of capacitors in RC circuits as in the Hodgkin-Huxley model \cite{hh, sch2}, subthreshold oscillations may be better modelled by LC circuits. The total energy of such a circuit is
\beq
U = \frac{1}{2}(CV^2 + LI^2) = \frac{1}{2}\left(\frac{C^2V^2}{C} + \frac{L^2I^2}{L}\right)  
\eeq
where $C$ is the capacitance, $L$ is the inductance, $V$ is the voltage and $I$ the current. This can be written in the form
\ben
H &=& \frac{1}{2}\left(\frac{L^2I^2}{L} +\frac{C^2V^2}{C}\right)\nonumber\\&:=& \frac{p^2}{2m} + \frac{\omega^2 Lx^2}{2} = \frac{p^2}{2m} + \frac{m\omega^2 x^2}{2}  
\een
with $U = H$ (the Hamiltonian), $L = m$ (the mass), $x = CV$ (the coordinate), $I = C dV/dt = dx/dt$ (the velocity), $p = LI$ (the momentum) and $\omega = 1/\sqrt{LC}$ (the angular frequency). The identical formal structure of a mechanical and LC harmonic oscillator prompts us to speculate that the quantum mechanical description of the LC oscillator is in the form of a state vector $|\psi\rangle$. Then, in a representation in which $x$ is an independent variable and $|\psi(x, t)|^2dx$ is the probability that the oscillator has coordinate between $x$ and $x + dx$, the time evolution of the state function is governed by the equation of motion
\beq
i\hat{\hbar} \partial_t |\psi\rangle = \hat{H}|\psi\rangle,\,\,\,\,\hat{H} = \frac{\hat{p}^2}{2m} + \frac{1}{2}m\omega^2\hat{x}^2 \label{ho}.
\eeq
This corresponds exactly to the Schr\"{o}dinger equation (\ref{sc2}) with $\psi(x,t) = \langle x|\psi\rangle$, $\hat{p} = -i\hat{\hbar}\partial_x$, $V = m\omega^2x^2/2$. We therefore conclude that the Schr\"{o}dinger equation (\ref{sc2}) with $V = m\omega^2x^2/2$ describes the subthreshold oscillations of neurons.

Since we are interested in membrane oscillations, we need time-dependent solutions of (\ref{sc2}) with $V = m\omega^2x^2/2$. Such states are called `coherent states' $|\alpha\rangle$ of the harmonic oscillator. They were first introduced by Schr\"{o}dinger in 1926, and their theory was developed further by Glauber \cite{gl} and Sudarshan \cite{sud} in connection with laser optics. Their importance lies in the fact that they are `minimum uncertainty' states and closely approximate classical oscillators. One defines the `displacement operator' $D(\alpha)$ as $D(\alpha) = e^{\alpha a^\dagger - \alpha^* a}$ where $\alpha = |\alpha|e^{i\phi}$ is a complex number. A coherent state $|\alpha\rangle$ is then defined as a displaced vacuum state $|0\rangle$, 
\ben
|\alpha\rangle &=& D(\alpha)|0\rangle,\\
a|0\rangle &=& 0.
\een
Using $\alpha(t) = \alpha e^{-i\omega t} = |\alpha|e^{-i(\omega t - \phi)}$, one can verify (see the Appendix for details) that the expectation value of $x$ in such a state is 
\beq
\langle x\rangle = \sqrt{2}x_0 |\alpha|\cos(\omega t - \phi)
\eeq
where $x_0 = \sqrt{\frac{\hat{\hbar}}{m\omega}}$ is the width of the harmonic oscillator ground state which is a gaussian distribution. The uncertainties in position and momentum in such a state satisfy the relation $(\sigma_x)_0^2(\sigma_p)_0^2 = \hat{\hbar}^2/4$, showing they are minimum uncertainty states, i.e. states with minimum quantum uncertainty and hence closest to classical states. 
\begin{figure}[H]
\centering
{\includegraphics[scale=0.8]{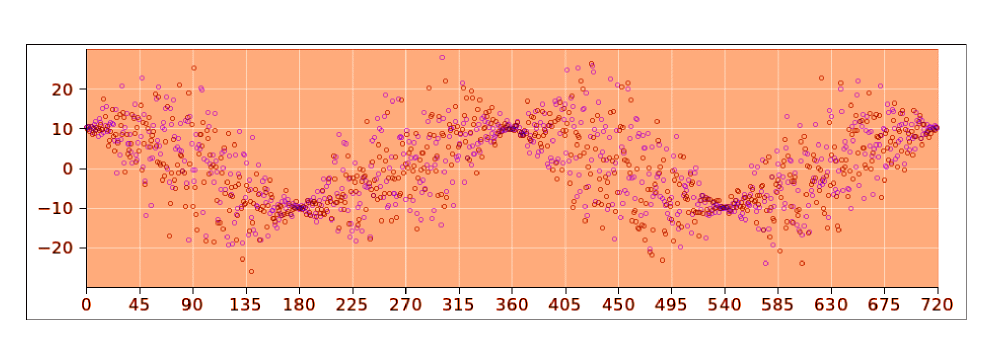}}
\caption{\label{Figure 5}{\footnotesize Coherent state representation of the harmonic oscillator: the expectation value $\langle x\rangle$ as a function of time in the ground state of the harmonic oscillator, showing minimum uncertainty scatter points around a classical cosine wave.}}
\end{figure}

Hence, the unequivocal prediction is that subthreshold neural oscillations should be sinusoidal with a scatter of $\hat{\hbar}^2/4$ around the classical values. A measurement of the scatter will therefore determine the value of $\hat{\hbar}$. Future work should therefore investigate quantum-like fluctuations in subthreshold neural oscillations using, for example, high-resolution electrophysiological recordings, such as patch-clamp techniques \cite{sak, lov}, capable of detecting minute fluctuations in membrane potentials.

Another possibility is to search for the discrete stationary energy levels $E_n = (n + \frac{1}{2})\hat{\hbar}\omega$ of neurons implied by the harmonic oscillator model (\ref{ho}) (see the Appendix). These are solutions of the time-independent Schr\"{o}dinger equation and should be observable during the quiescent or refractory periods of the neurons. The Helmholtz free energy $F$ and the average energy $\langle E\rangle$ of a quantum harmonic oscillator are given by (see the Appendix)
\ben
F &=& \frac{\hat{\hbar}\omega}{2} + \frac{1}{\beta}\ln \left(1 - e^{-\beta \hat{\hbar}\omega}\right),\\
\langle E\rangle &=& \frac{\hat{\hbar}\omega}{2}  + \frac{\hat{\hbar}\omega e^{-\hat{\hbar}\omega}}{1 - e^{-\beta \hat{\hbar}\omega}}
\een
where $\beta = 1/kT$, $T$ being the absolute temperature. These expressions show that both $F$ and $\langle E\rangle$ tend to the zero-point energy as the temperature $T$ tends to zero. These results are experimentally verifiable in principle and can be used to determine the value of $\hat{\hbar}$.
 
Before concluding, it should be mentioned that `entangled states' also exist in stochastic mechanics, and stochastic mechanics and quantum mechanics agree in predicting all observed correlations at different times. The reader is referred to the papers by Faris \cite{far} and Petroni and Morato \cite{pet} for details. Looking for entanglement in neural systems is therefore another important area for further research, not only for its intrinsic value but also because entanglement is a key resource in quantum information processing, and it is important to find out whether the brain makes use of it, as conjectured \cite{khren}. We predict that the presence of non-classical correlations in neuronal fluctuations, akin to quantum entanglement, could be detected through cross-correlation analyses of simultaneous recordings from neighbouring neurons.

Finally, given that neural plasticity is often linked to the probabilistic nature of synaptic changes, our results suggest that quantum-like effects might influence how plasticity occurs in specific neural circuits. These insights could open new avenues for understanding cognitive processes, where classical and quantum descriptions might intersect, leading to a more comprehensive theory of brain function.

\section{Appendix}
When considering harmonic oscillators, it is convenient to introduce the ladder operators
\ben
a &=& \frac{1}{\sqrt{2m\omega\hat{\hbar}}}(m\omega x + i\hat{p}),\\
a^\dagger &=& \frac{1}{\sqrt{2m\omega\hat{\hbar}}}(m\omega x - i\hat{p}).
\een
Using the commutation rule $[\hat{p}, x]= -i\hat{\hbar}$, one gets the commutation relation $[a,a^\dagger] = 1$, and 
\ben
a^\dagger |n\rangle &=& \sqrt{n+1}|n+1\rangle,\\
a|n\rangle &=& \sqrt{n}|n-1\rangle,\\
a^\dagger a|n\rangle &=& n|n\rangle
\een
where $n = 0,1,2,...$ is an integer.
The Hamiltonian operator can be written in the form
\beq
\hat{H} = \hat{\hbar}\omega\left(a^\dagger a + \frac{1}{2}\right) = \hat{\hbar}\omega\left(N + \frac{1}{2}\right)  
\eeq
where $N = a^\dagger a$ is the number operator. It then follows from the time-independent Schr\"{o}dinger equation 
$\hat{H}\psi_n = E_n\psi_n$ that the energy eigenvalues are 
\beq
E_n = \hat{\hbar}\omega\left(n + \frac{1}{2}\right).
\eeq The factor $\hat{\hbar}\omega/2$ is the `zero-point energy' of neurons. The corresponding energy eigenfunctions are given by
\beq
\psi_n(x) = \frac{1}{\sqrt{n!}}(a^\dagger)^n\psi_0(x) 
\eeq
with the ground state
\beq
\psi_0(x) = \left(\frac{m\omega}{\pi\hat{\hat{\hbar}}}\right)e^{-\frac{m\omega x^2}{2\hat{\hat{\hbar}}}}
\eeq
which is a Gaussian distribution with width $x_0 = \sqrt{\frac{\hat{\hbar}}{m\omega}}$. 

Now note that the coherent state can be written as
\ben
a|\alpha\rangle &=& \alpha|\alpha\rangle,\\
\langle \alpha|a^\dagger &=& \langle \alpha|\alpha^*,\\
\langle \alpha|a^\dagger a|\alpha\rangle &=& |\alpha|^2.
\een
Using these results, one can compute the uncertainties in $x$ and $p$:
\ben
(\sigma_x)^2 &=& \langle x^2\rangle - \langle x\rangle^2 = x_0^2 \left(n + \frac{1}{2}\right),\\  
(\sigma_p)^2 &=& \langle p^2\rangle - \langle p\rangle^2 = \frac{\hat{\hbar}^2}{x_0^2} \left(n + \frac{1}{2}\right),
\een
and hence for the ground state ($n=0$) $(\sigma_x)_0^2(\sigma_p)_0^2 = \hat{\hbar}^2/4$. For further details of coherent states the reader is referred to Ref \cite{bert}.

The thermodynamic properties of the quantum harmonic oscillator can be calculated using the standard techniques of statistical mechanics. The partition function 
\beq
Z= Tr e^{-\beta\hat{H}} = \sum_{n=0}^\infty e^{-\beta E_n} = \sum_{n=0}^\infty e^{-\beta \left(n + \frac{1}{2}\right)\hat{\hbar}\omega} = \frac{1}{2} csch \left(\frac{\beta \hat{\hbar}\omega}{2}\right)
\eeq
where $\hat{H}$ is the Hamiltonian operator. Hence
\ben
F &=& -\frac{1}{\beta}\ln Z = \frac{\hat{\hbar}\omega}{2} + \frac{1}{\beta}\ln \left(1 - e^{-\beta \hat{\hbar}\omega}\right),\\
\langle E\rangle &=& - \frac{\partial \ln Z}{\partial \beta} = \frac{\hat{\hbar}\omega}{2}  + \frac{\hat{\hbar}\omega e^{-\hat{\hbar}\omega}}{1 - e^{-\beta \hat{\hbar}\omega}}.
\een

\end{document}